\documentclass[twocolumn]{aastex6}
\usepackage{graphicx}
\usepackage{amssymb,amsfonts,amsmath,amstext,amsgen,amsopn,amsxtra,indentfirst,times,longtable}

\begin{document}

\title{Jupiter as an Exoplanet: Insights from Cassini Phase Curves}

\author{Kevin Heng\altaffilmark{1,2}}
\author{Liming Li\altaffilmark{3}}
\altaffiltext{1}{University of Bern, Center for Space and Habitability, Gesellschaftsstrasse 6, CH-3012, Bern, Switzerland.  Email: kevin.heng@csh.unibe.ch}
\altaffiltext{2}{University of Warwick, Department of Physics, Astronomy \& Astrophysics Group, Coventry CV4 7AL, United Kingdom. Email: Kevin.Heng@warwick.ac.uk}
\altaffiltext{3}{University of Houston, Department of Physics, Houston, TX 77004, U.S.A. Email: lli13@uh.edu}

\begin{abstract}
Due to its proximity to Earth, Jupiter of the Solar System serves as a unique case study for gas-giant exoplanets.  In the current Letter, we perform fits of ab initio, reflective, semi-infinite, homogeneous model atmospheres to 61 phase curves from 0.40 to 1.00 $\mu$m, obtained from the Cassini spacecraft, within a Bayesian framework.  We reproduce the previous finding that atmospheric models using classic reflection laws (Lambertian, Rayleigh, single Henyey-Greenstein) provide poor fits to the data.  Using the double Henyey-Greenstein reflection law, we extract posterior distributions of the single-scattering albedo and scattering asymmetry factors and tabulate their median values and uncertainties.  We infer that the aerosols in the Jovian atmosphere are large, irregular, polydisperse particles that produce strong forward scattering together with a narrow backscattering lobe.  The near-unity values of the single-scattering albedos imply that multiple scattering of radiation is an important effect.  We speculate that the observed narrow backscattering lobe is caused by coherent backscattering of radiation, which is usually associated with Solar System bodies with solid surfaces and regolith.  Our findings demonstrate that precise, multi-wavelength phase curves encode valuable information on the fundamental properties of cloud/haze particles.  The method described in this Letter enables single-scattering albedos and scattering asymmetry factors to be retrieved from James Webb Space Telescope phase curves of exoplanets.
\end{abstract}


\section{Introduction}
\label{sect:intro}

Named after the Roman god of the sky and thunder, Jupiter is the most massive planet of our Solar System, has been studied for several centuries and was the subject of close scrutiny by recent space missions \citep{porco04,bolton17}.  Since gas giants were the first exoplanets to be discovered \citep{mq95} and their atmospheres remain the most well-characterised \citep{ds17}, Jupiter holds a special place among Solar System planets as a unique case study for exoplanets and their atmospheres.

\begin{figure}
\begin{center}
\vspace{-0.2in}
\includegraphics[width=\columnwidth]{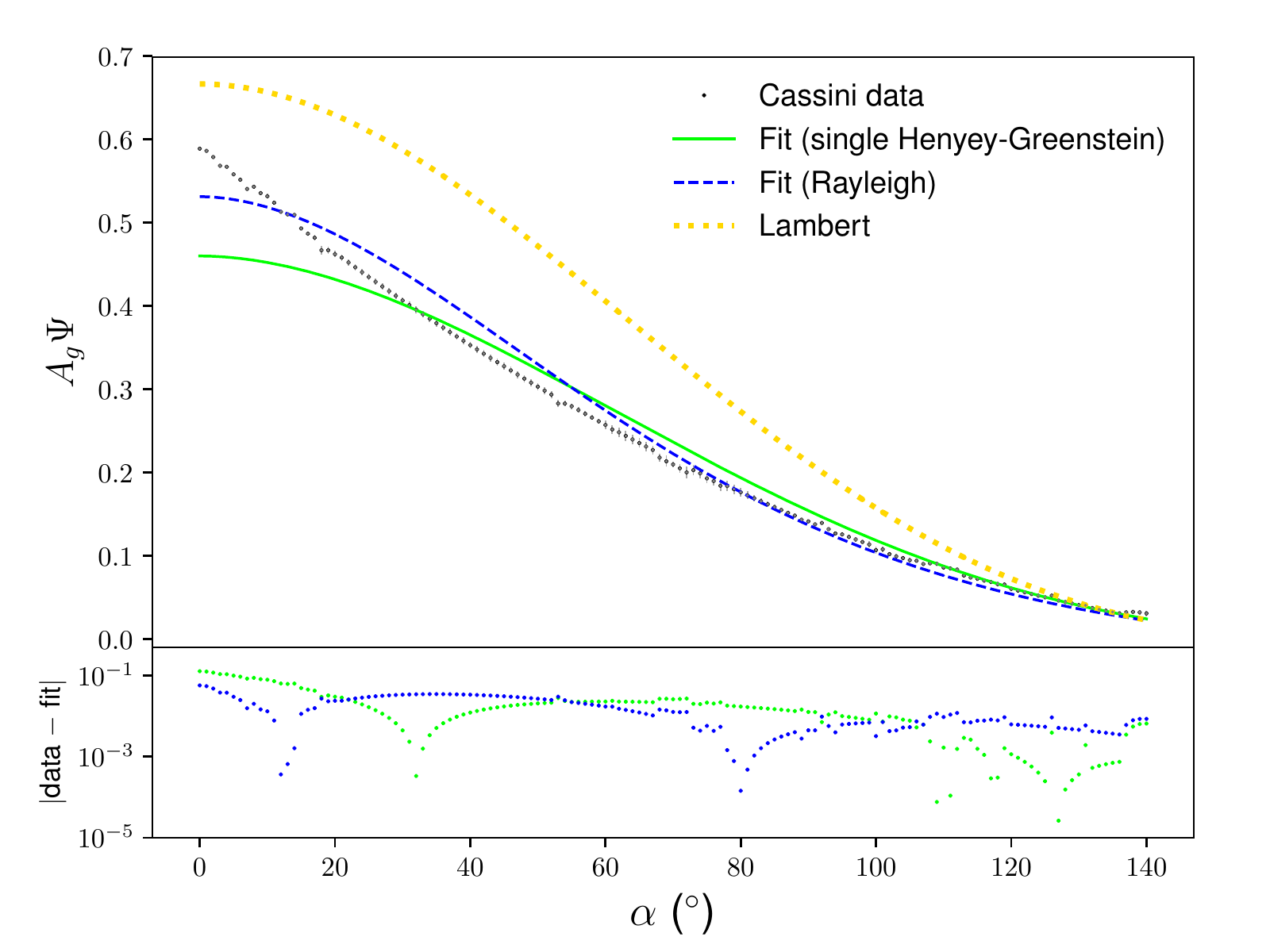}
\end{center}
\vspace{-0.2in}
\caption{Examples of failed fits of model phase curves of semi-infinite, homogeneous atmospheres to the 0.50 $\mu$m Cassini phase curve of Jupiter assuming the single Henyey-Greenstein and Rayleigh reflection laws.  The Lambertian reflection law is not a fit as it does not possess any free parameters (i.e., $\omega=1$ and $A_g=2/3$).}
\vspace{-0.15in}
\label{fig:failed_fits}
\end{figure}

The visible and near-infrared phase curves of Jupiter are of particular interest, because they quantify the fraction of sunlight reflected by the Jovian atmosphere as a function of the orbital phase angle.  Unlike for $\sim 1000$ K hot Jupiters, the Jovian phase curves are not contaminated by the thermal (infrared) emission of Jupiter and can be safely assumed to comprise predominantly reflected sunlight.  Using data from the Cassini space mission, \cite{dyu16} and \cite{mayorga16} previously showed that the Jovian phase curves are ``cuspy" and peak more sharply towards zero phase angle than classic laws of reflection (Lambertian, Rayleigh).  Figure \ref{fig:failed_fits} shows an example, where we reproduce the findings of \cite{dyu16} and \cite{mayorga16}.  Neither study performed fits for fundamental physical parameters (single-scattering albedo, scattering asymmetry factor) within a Bayesian framework, as such numerical calculations are computationally expensive.  

Recently, \cite{heng21} reported first-principles, fully analytical solutions of reflected light phase curves for any reflection law.  In the current Letter, we exploit this development to fit ab initio\footnote{There are no ``tuning parameters" beyond one's choice of the reflection law (scattering phase function $P$), since the formulae for $A_g$ and $\Psi$ are formal solutions of the radiative transfer equation \citep{heng21}.  The absence of parameters to finetune implies the ability to extract or retrieve fundamental physical parameters associated with the scatterers, which may take the form of atoms, molecules, ions or aerosols (clouds/hazes).} models of semi-infinite, homogeneous atmospheres to a set of 61 Cassini phase curves of Jupiter \citep{li18} from wavelengths of $\lambda=0.40$--1.00 $\mu$m (with bin sizes of 0.01 $\mu$m) within a Bayesian framework \citep{fm13}.  We demonstrate that the double Henyey-Greenstein (DHG) reflection law \citep{hg,k75,zl16} provides reasonable fits to the Jovian phase curves, which indicates the scattering of light by large, irregular particles \citep{mh95} that are poorly described by Mie theory \citep{mie,kh18}.

In Section \ref{sect:methods}, we describe our methods and data.  In Section \ref{sect:results}, we demonstrate that the set of fits yields inferred values (with uncertainties) of the single-scattering albedo and scattering asymmetry factors at each wavelength (Table 1), which provide important input for workers studying the chemistry of Jovian clouds and hazes.  In Section \ref{sect:discussion}, we compare our work to previous studies and discuss its implications.

\begin{figure*}
\begin{center}
\includegraphics[width=\columnwidth]{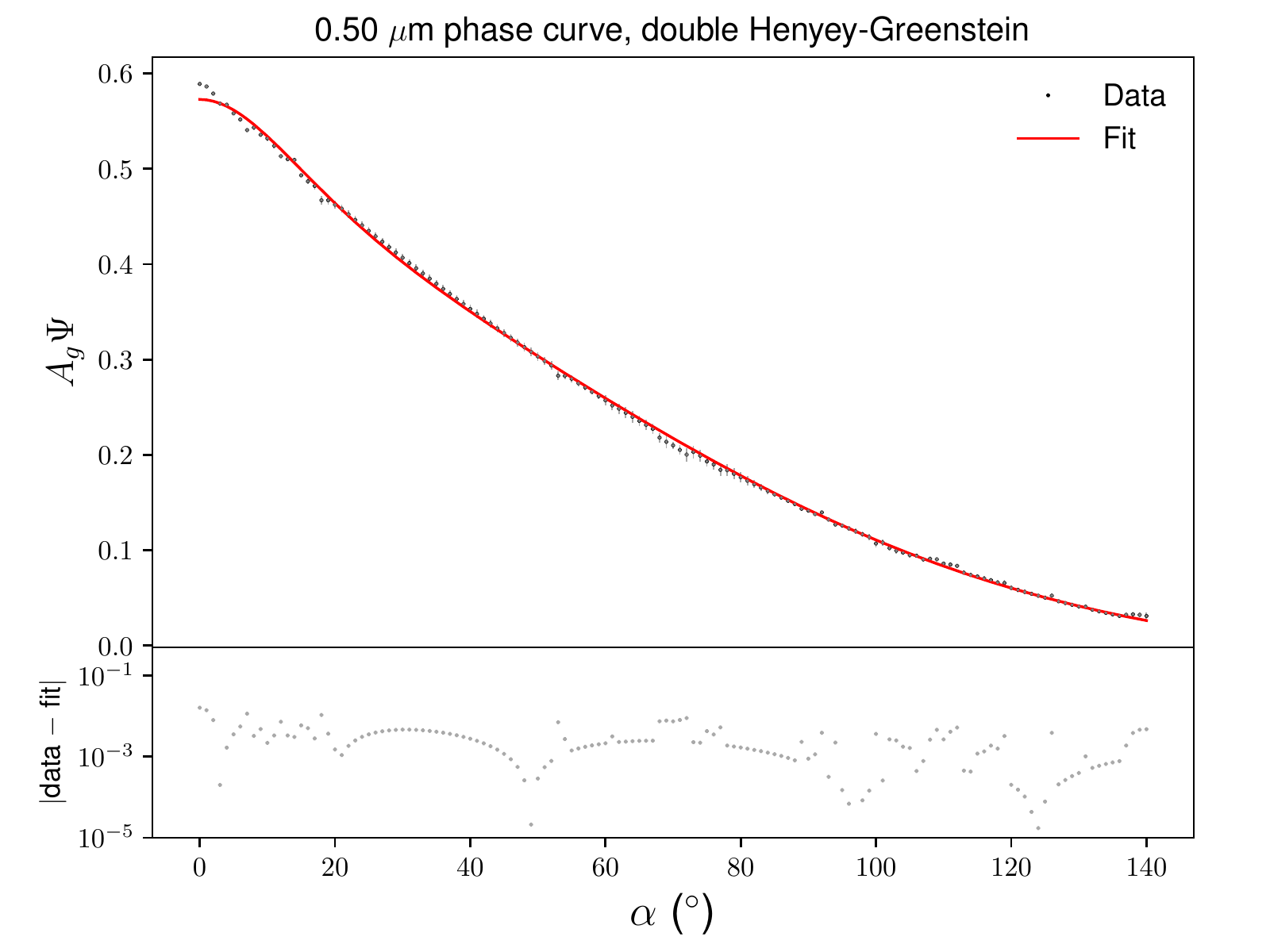}
\includegraphics[width=0.75\columnwidth]{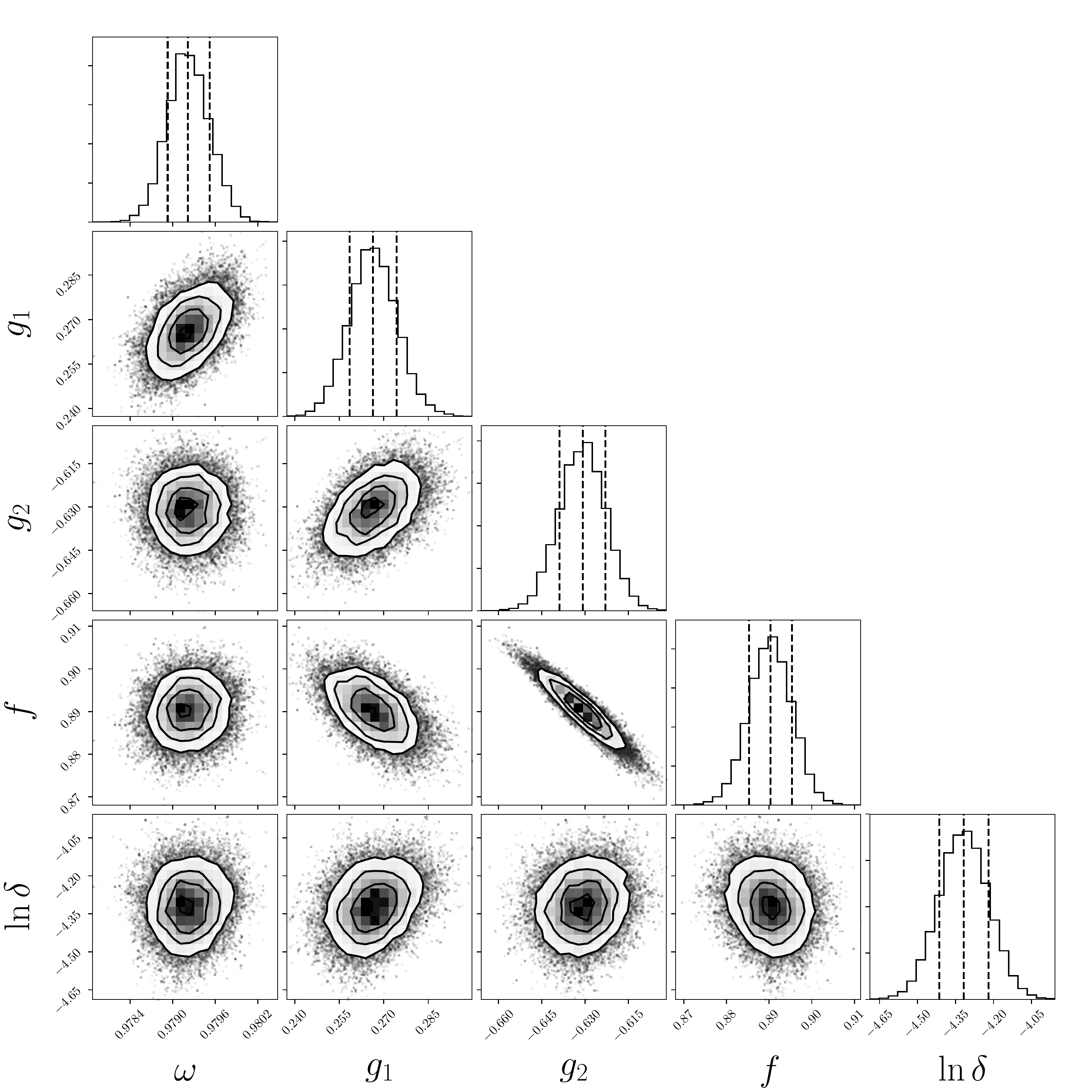}
\includegraphics[width=\columnwidth]{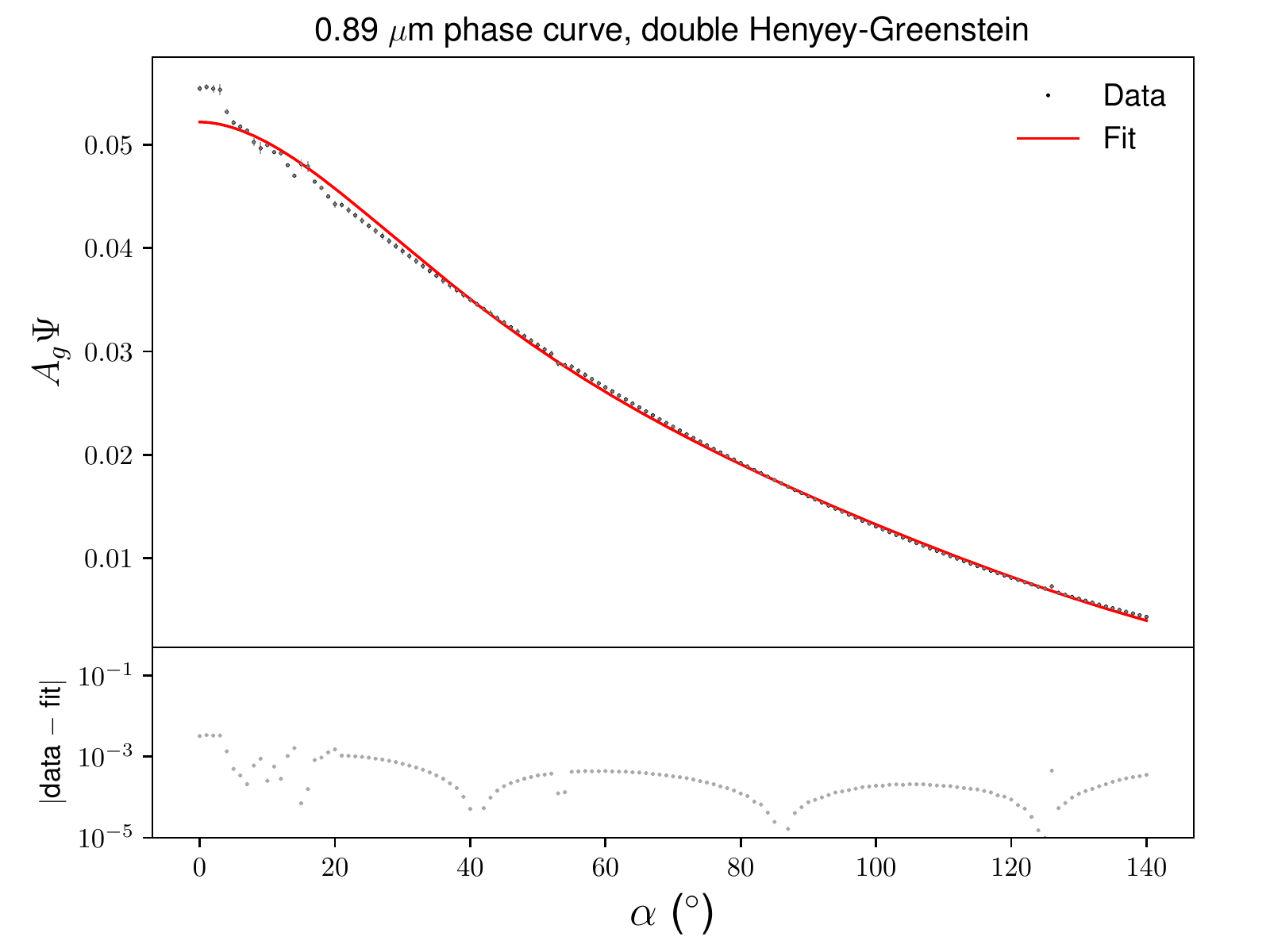}
\includegraphics[width=0.75\columnwidth]{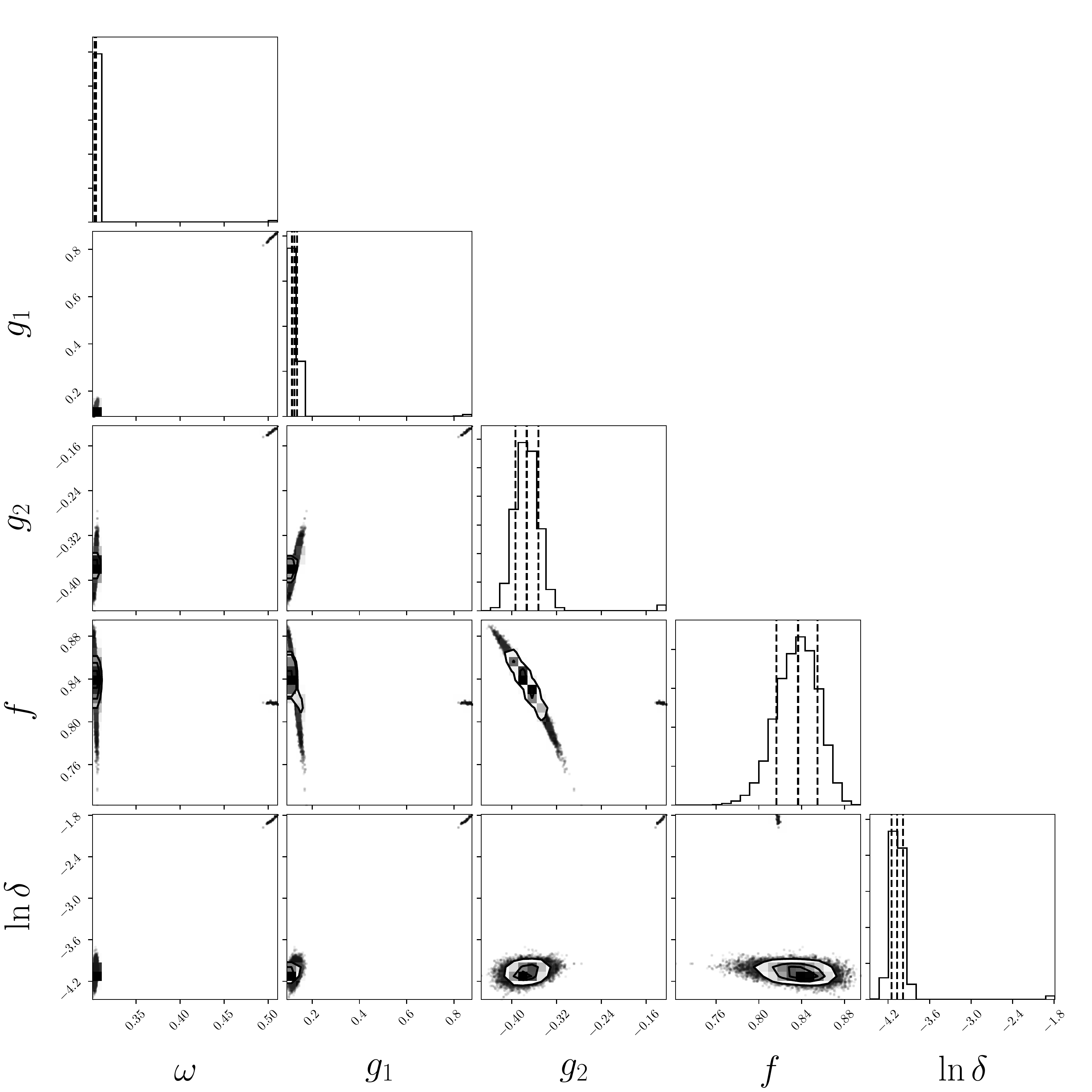}
\end{center}
\vspace{-0.2in}
\caption{Examples of fits to Cassini data at the peak of the solar spectrum (0.50 $\mu$m; top row) and in the MT3 methane absorption band (0.89 $\mu$m; bottom row).  The left column shows the data (and their uncertainties), fits and residuals, while the right column shows the corresponding posterior distributions of the DHG parameters.}
\label{fig:fits}
\end{figure*}

\section{Methods \& Data}
\label{sect:methods}

\subsection{Background theory}

Building on the classic work of \cite{chandra}, \cite{sobolev} and \cite{hapke81}, \cite{heng21} derived novel analytical solutions for the geometric albedo $A_g$ and integral phase function $\Psi$ of semi-infinite atmospheres for any law of reflection (scattering phase function $P$).  For context, some of the key findings of \cite{heng21} are concisely reviewed here.

Let incident sunlight with intensity $I_\star$ impinge upon an atmosphere at a zenith angle $\mu_\star$.  Let the intensity of light reflected by the atmosphere be $I_0$, which is generally a formal solution of the radiative transfer equation \citep{chandra,hapke81}.  The reflection coefficient is \citep{dy74,sobolev,seager}
\begin{equation}
\rho = \frac{I_0}{I_\star \mu_\star}.
\end{equation}
The quantity $\rho/\pi$ is sometimes termed the ``bidirectional reflection distribution function" (BRDF; e.g., \citealt{dyu16}).

Define $\rho_0 \equiv \rho(\alpha=0^\circ)$.  The geometric albedo may be expressed in terms of the reflection coefficient \citep{dy74,sobolev},
\begin{equation}
A_g = 2 \int^1_0 \rho_0 \mu^2 ~d\mu,
\end{equation}
where $\mu = \cos\theta$ and $\theta$ is the polar angle in the frame of reference of the planet.  In the observer-centric coordinate system, the flux (divided by $I_\star$) observed at any orbital phase angle $\alpha$ is \citep{sobolev},
\begin{equation}
F = \int^{\pi/2}_{\alpha-\pi/2} \int^{\pi/2}_0 \rho \mu \mu_\star ~\cos\Theta ~d\Theta ~d\Phi,
\end{equation}
where $\Theta$ and $\Phi$ are the observer's latitude and longitude, respectively.   If we define $F_0 \equiv F(\alpha=0^\circ)$, then the integral phase function is \citep{sobolev}
\begin{equation}
\Psi = \frac{F}{F_0}.
\end{equation}

The approach of \cite{hapke81} is followed in assuming that single scattering is described by any scattering phase function $P$, but multiple scattering occurs isotropically.  Building on this rich body of work, \cite{heng21} showed that closed-form solutions for $A_g$ and $\Psi$ may be derived for \textit{any} $P$.  Within this formalism, the phase curve is $A_g \Psi$.  In other words, both the shape \textit{and} normalisation of the reflected light phase curve are computed from first principles.

\begin{figure*}
\begin{center}
\includegraphics[width=\columnwidth]{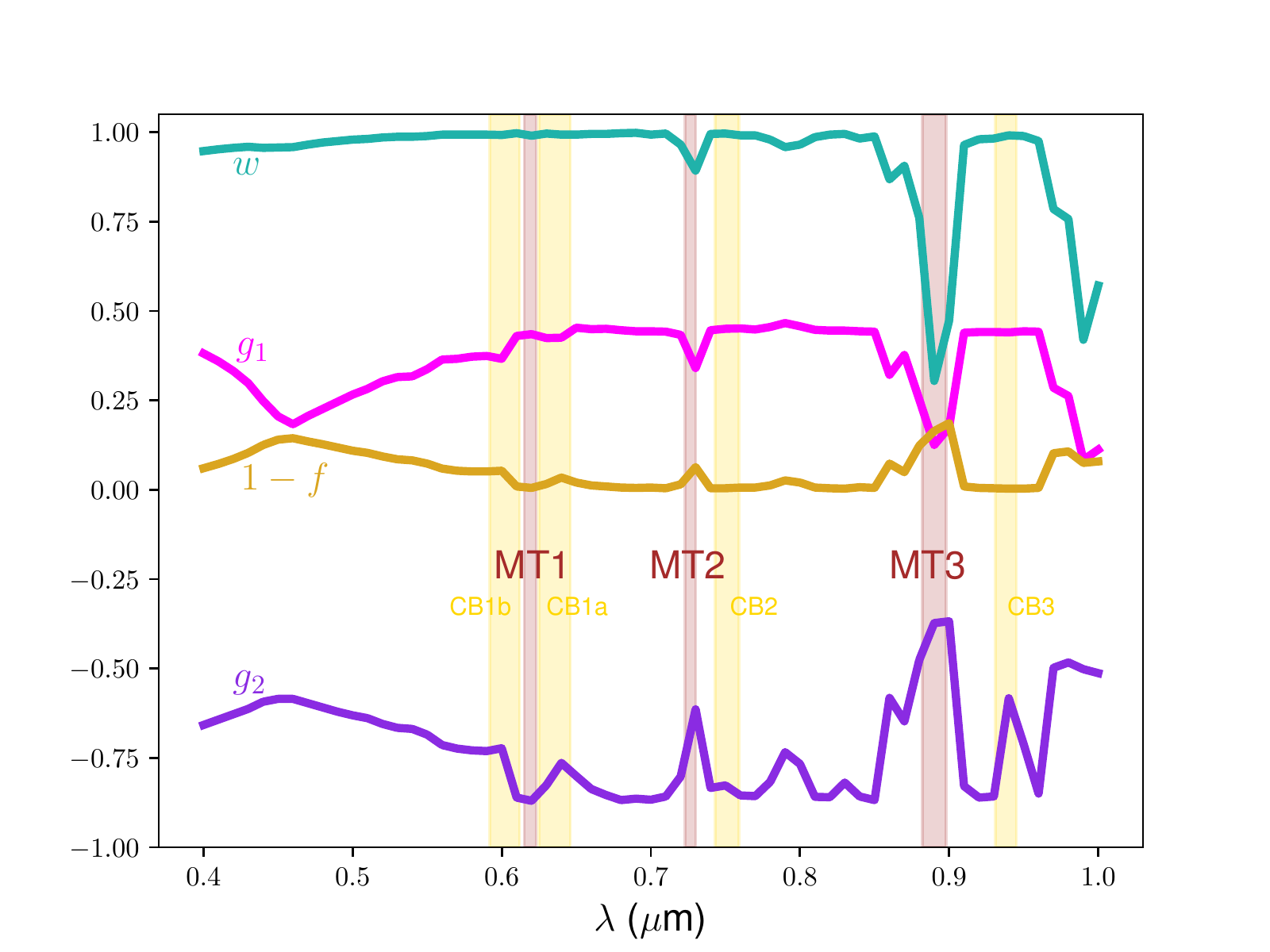}
\includegraphics[width=\columnwidth]{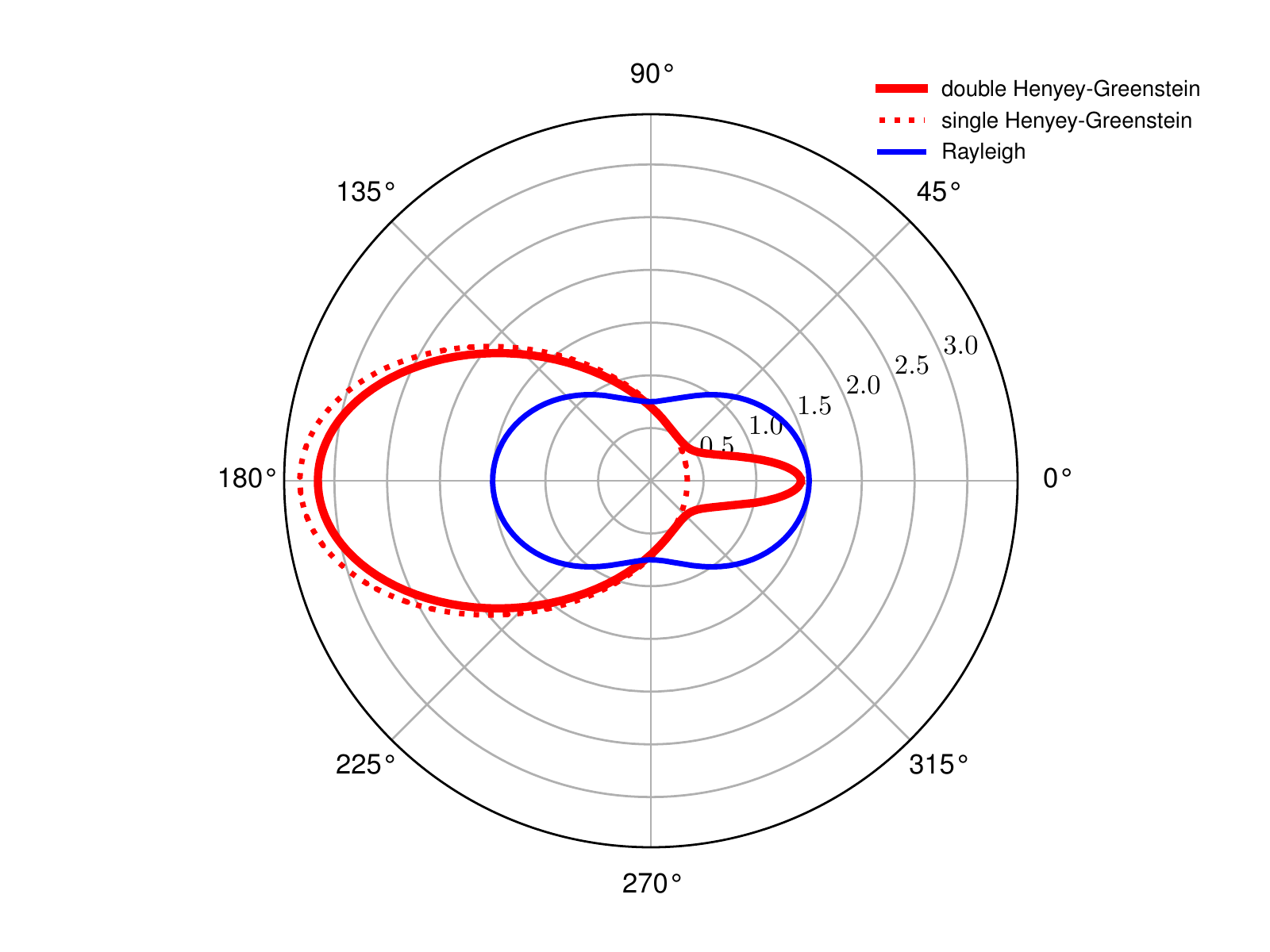}
\end{center}
\vspace{-0.2in}
\caption{The left panel shows the inferred DHG parameters: single-scattering albedo $\omega$, scattering asymmetry factors $g_1$ and $g_2$ and weighting factor $1-f$.  Inferred uncertainties on the parameters are not shown as they are typically smaller than the width of the line.  The right panel shows visualisations of the single Henyey-Greenstein (adopting a mean value of $\bar{g_1}=0.36$), double Henyey-Greenstein (adopting mean values of $\bar{g_1}=0.36$, $\bar{g_2}=-0.71$ and $\bar{f}=0.95$) and Rayleigh scattering phase functions.  Isotropic scattering is represented by a unit circle (not shown).  Angles stated are the values of $\alpha=180^\circ-\beta$ with $\beta$ being the scattering angle.  Generally, the particles simultaneously exhibit strong forward scattering ($f \sim 1$) and a narrow backscattering lobe.  The MT1 (0.615--0.623 $\mu$m), MT2 (0.723--0.730 $\mu$m) and MT3 (0.882--0.898 $\mu$m) methane absorption bands \citep{porco04} are indicated with translucent bars.  The corresponding continuum bands are CT1b (0.592--0.612 $\mu$m), CT1a (0.625--0.646 $\mu$m), CT2 (0.743--0.759 $\mu$m) and CT3 (0.931--0.945 $\mu$m), which are also indicated with translucent bars.}
\label{fig:DHG}
\end{figure*}

\subsection{Double Henyey-Greenstein reflection law}

The DHG scattering phase function uses a pair of Henyey-Greenstein scattering phase functions\footnote{The terms ``scattering phase function" and ``reflection law" are used synonymously.} \citep{hg} to describe the presence of both forward and reverse scattering.  It is given by \citep{k75,zl16}
\begin{equation}
P = \frac{f \left(1 - g_1^2\right)}{\left( 1 + g_1^2 + 2 g_1 \cos \alpha \right)^{3/2}} + \frac{\left(1-f\right) \left(1 - g_2^2\right)}{\left( 1 + g_2^2 + 2 g_2 \cos \alpha \right)^{3/2}}.
\end{equation}
The two scattering asymmetry factors $0 \le g_1 \le 1$ and $-1 \le g_2 \le 0$ quantify the degree of forward and reverse scattering, respectively.  The factor $0 \le f \le 1$ quantifies the relative importance of each Henyey-Greenstein scattering phase function.  Neither the single nor the double Henyey-Greenstein scattering phase functions are derived from first principles, but are rather ad hoc functions that possess convenient mathematical properties associated with Legendre polynomials (e.g., \citealt{sobolev}).

\subsection{Bayesian framework for data fitting}

To compute $A_g \Psi$ requires the specification of 4 free parameters: $g_1$, $g_2$, $f$ and the single-scattering albedo $\omega$.  We perform fits of $A_g \Psi$ to the Jovian phase curves with the open-source Markov Chain Monte Carlo code \texttt{emcee} \citep{fm13} using 100 walkers and 1000 steps with a burn-in of 500 steps\footnote{On a 2016 MacBook Pro with a 2.9 GHz Intel Core i7, each fit took about 12--15 seconds.}.  A Gaussian likelihood function is assumed.  Within the likelihood function, we account for the possibility that the data uncertainties have been under-estimated by adding $(\delta A_g \Psi)^2$ to the variance, where $\ln{\delta}$ is formally part of the fit \citep{hogg10}\footnote{\scriptsize \texttt{https://emcee.readthedocs.io/en/stable/tutorials/line/}}.  Uniform prior distributions are assumed: $0 \le \omega \le 1$, $0 \le g_1 \le 1$, $-1 \le g_2 \le 0$, $0 \le f \le 1$ and $-10 < \ln{\delta} < 1$.  The joint posterior distributions of the parameters are plotted using the \texttt{corner} routine written in the Python programming language \citep{fm16}.  

\subsection{Cassini data of Jupiter}

The global images of Jupiter recorded by the Cassini spacecraft are used to generate the light curves at visible and near-infrared wavelengths \citep{li18}.  The Cassini spacecraft observed Jupiter from October 2000 to March 2001 en route to Saturn.  Among the dozen scientific instruments of Cassini, the Imaging Science Subsystem (ISS; \citealt{porco04}) and the Visual and Infrared Mapping Spectrometer (VIMS; \citealt{brown04}) conducted observations of Jupiter at visible and near-infrared wavelengths. The ISS is an imager with two cameras and multiple filters onboard the Cassini spacecraft \citep{porco04, li18}.  We mainly use the ISS observations because of their better spatial resolution and more complete coverage of phase angle compared to the VIMS observations. To our knowledge, the Cassini ISS provides the best coverage of phase angles among all available global images of Jupiter.  The selected high-spatial-resolution ISS global images recorded have a range of phase angles from about $0^\circ$ to $140^\circ$.  This limit of $140^\circ$ comes from the need to prevent stray light from entering the camera while pointing close to the Sun.  The ISS global images have observational gaps in phase angle, which are filled by a polynomial function \citep{li18} using a least-squares method \citep{br03}.  In addition, the ISS global images were recorded at discrete wavelengths and some observational gaps in wavelength exist.  We interpolate the phase curves from the ISS wavelengths to all wavelengths from 0.40--1.00 $\mu$m (0.01 $\mu$m bin size) by using ground-based spectra measured at a phase angle of $6.8^\circ$ \citep{k98}.  Each phase curve has 141 data points ($\alpha = 0^\circ$--140$^\circ$ with $1^\circ$ resolution).

There are two dominant sources of uncertainties in our processing of Jupiter's light curves: 1. Calibrating the Cassini ISS global images; 2. Filling in the observational gaps in phase angle and wavelength.  Calibration was performed using the Cassini ISS CALibration (\texttt{CISSCAL}) software (\texttt{http://ciclops.org/sci/cisscal.php}).  The calibration uncertainties (i.e., uneven bit-weighting, bias subtraction, 2-Hz noise, dark current in the ISS cameras, bright/dark pixel pair artifacts from anti-blooming mode, flat-field artifacts) are discussed in detail in \cite{west10} and \cite{k20}.  All of these uncertainties typically amount to a few percent of the calibrated radiance \citep{k20}.  We apply the correction factors provided by the ISS calibration team to account for the aforementioned calibration uncertainties.  The remaining calibration uncertainties should be less than one percent of the calibrated radiance.  To account for the uncertainties associated with filling in the observational gaps in phase angle and wavelength, we use the fit residuals, which are the differences between the observed and fitted values at the ISS observed phase angles and wavelengths, to estimate the uncertainties associated with filling in the observational gaps.  By fitting for $\ln{\delta}$, we formally allow for the possibility that the uncertainties have been under-estimated \citep{hogg10}.


\section{Results}
\label{sect:results}

\begin{figure*}
\begin{center}
\vspace{-0.2in}
\includegraphics[width=\columnwidth]{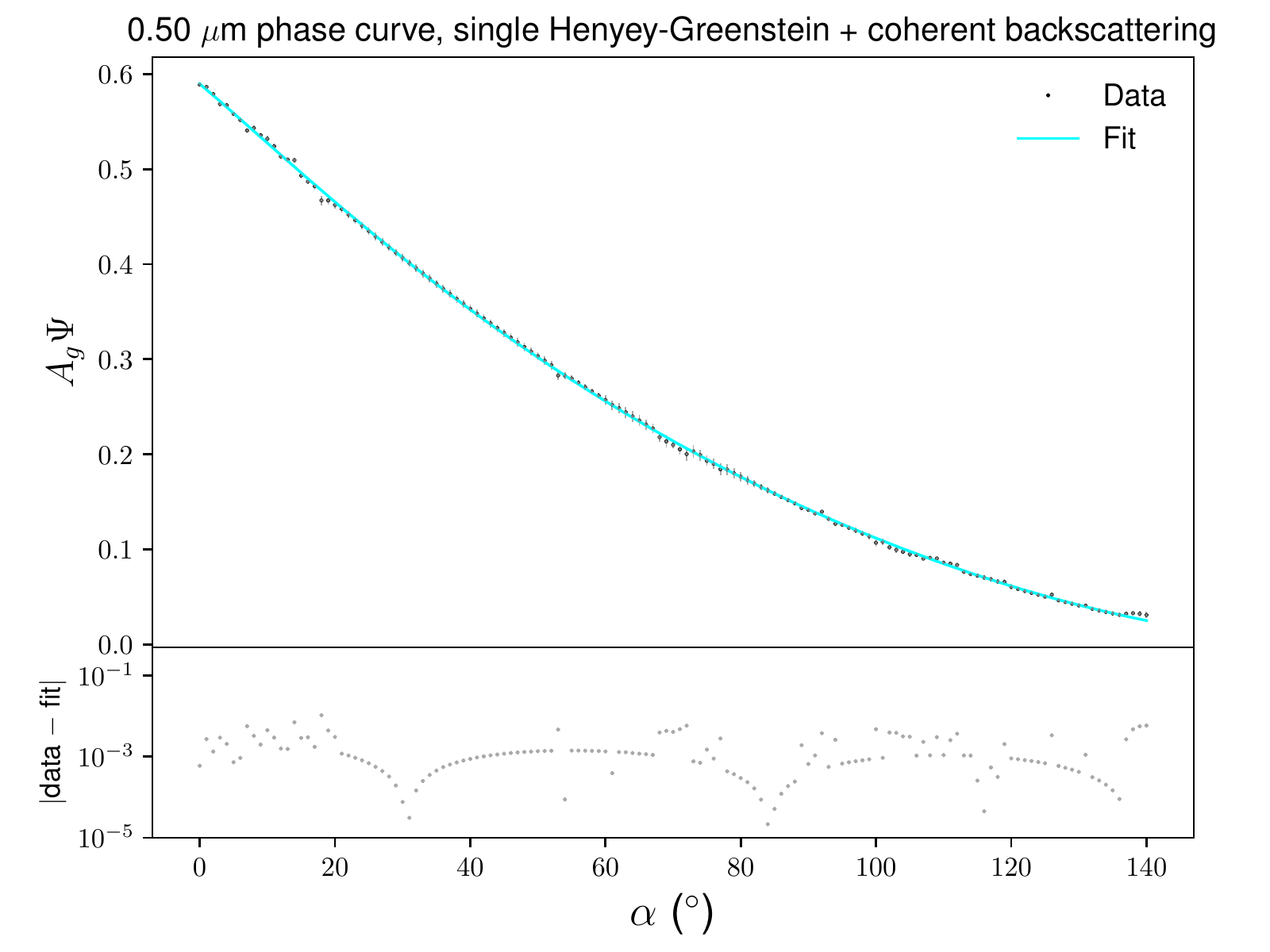}
\includegraphics[width=0.75\columnwidth]{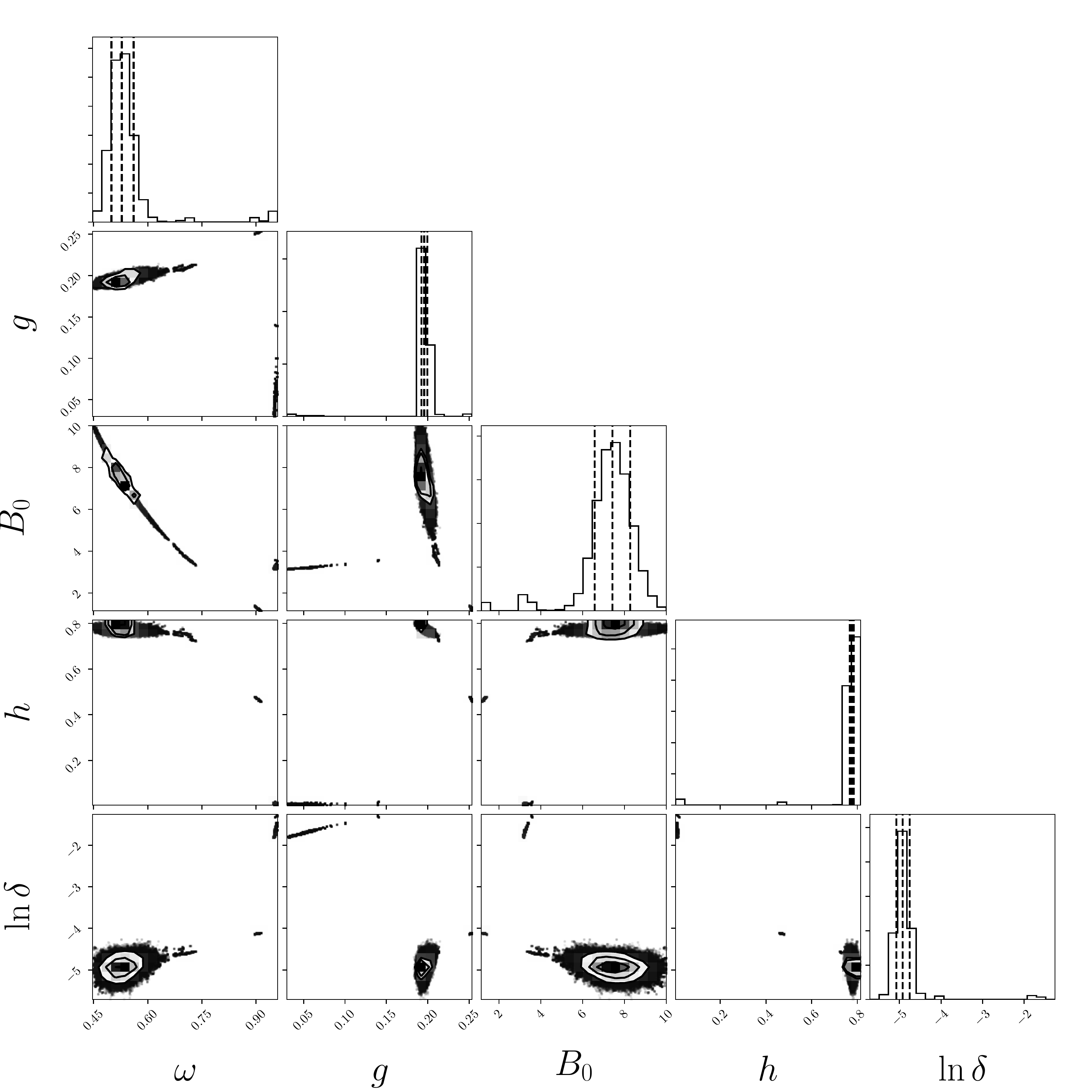}
\end{center}
\vspace{-0.2in}
\caption{Fit to the 0.50 $\mu$m phase curve using the single Henyey-Greenstein reflection law and a simplified function for the coherent backscattering of radiation.  The retrieved parameter values are $\omega=0.528^{+0.033}_{-0.029}$, $g=0.196^{+0.004}_{-0.003}$, $B_0=7.43^{+0.84}_{-0.86}$, $h=0.776^{+0.008}_{-0.009}$ and $\ln{\delta} = -4.92^{+0.17}_{-0.16}$.  Uniform prior distributions were adopted: $0 \le \omega \le 1$, $-1 \le g \le 1$, $0 \le B_0 \le 10$, $0 \le h \le 10$ and $-10 \le \ln{\delta} \le 1$. As there are hints of bimodality in some of the posterior distributions, we performed this particular fit for 10,000 Monte Carlo steps.}
\vspace{-0.05in}
\label{fig:CB}
\end{figure*}

\subsection{Reproducing the inadequacy of classic reflection laws}

Figure \ref{fig:failed_fits} reproduces the phenomenon already noted by \cite{dyu16}, \cite{mayorga16} and \cite{li18}, albeit within a Bayesian framework: the Jovian phase curves are generally too cuspy near $\alpha=0^\circ$ and are poorly fitted by the single Henyey-Greenstein and Rayleigh reflection laws.  By definition, the Lambertian reflection law always has $\omega=1$ and $A_g=2/3$ \citep{sobolev}; it possesses no free parameters and a fit is not performed.

\subsection{Fitting for fundamental physical parameters}

In Table 1, the median values of $\omega$, $g_1$, $g_2$, $f$ and $\ln{\delta}$, along with their 1-$\sigma$ uncertainties, are reported.  Figure \ref{fig:fits} shows examples of fits to data at 0.50 $\mu$m (the peak of the solar spectrum) and 0.89 $\mu$m (the MT3 methane absorption band; Table VIII of \citealt{porco04}).  Generally, the reduced chi-square value of the fits is always much less than unity.  Typically, we have $\delta \lesssim e^{-4} \approx 2\%$, but in some cases we have $\delta \sim e^{-2.3} \approx 10\%$.

Figure \ref{fig:DHG} shows the retrieved values of $\omega$, $g_1$, $g_2$ and $1-f$ as functions of wavelength.  It is striking that sharp decreases in $\omega$ and $g_1$, as well as an increase in $g_2$ (which is a negative quantity), coincide with the MT2 (0.723--0.730 $\mu$m) and MT3 (0.882--0.898 $\mu$m) methane absorption filters.  This behavior is consistent with a reduction in the strength of scattering.  It is less noticeable for the MT1 (0.615--0.623 $\mu$m), but it has been previously noted that methane absorption in this filter is weak \citep{li06}.

Over the wavelength range considered (0.40--1.00 $\mu$m), the mean values of the DHG parameters are 
\begin{equation}
\bar{\omega} \approx 0.93, ~\bar{g_1} \approx 0.36, ~\bar{g_2} \approx -0.71, ~\bar{f} \approx 0.95.
\end{equation}
Figure \ref{fig:DHG} shows a visualisation of the average scattering phase function, which is inconsistent with Rayleigh scattering.  The particles exhibit strong forward scattering ($f \sim 1$) together with a narrow backscattering lobe.  In other words, most of the power is in the Henyey-Greenstein scattering phase function that quantifies forward scattering.

\subsection{Interpretation}

A detailed interpretation of the inferred values of the DHG parameters in the context of scattering theory and atmospheric chemistry is beyond the scope of the current Letter, since the particles are likely to be non-spherical and chemically heterogeneous; see, e.g., \cite{zhang13} and \cite{guerlet20} for recent discussions.  However, several general statements may be made.  Firstly, the general inference of $f \sim 1$ implies that the narrow backscattering lobe is a higher-order correction to a single Henyey-Greenstein reflection law describing large particles that produce forward scattering.  Secondly, the general inference that $g_1>0$ indicates the presence of large particles with sizes that are somewhat larger than the range of wavelengths probed.  Thirdly, the non-monotonic variation of $g_1$ with wavelength indicates the presence of a size distribution---the particles are polydisperse.  If the particles were monodisperse, then $g_1$ would monotonically decrease with increasing wavelength.  Fourthly, the high inferred values of $\omega \sim 1$ indicate that multiple scattering of sunlight is a significant effect \citep{hapke81} and are consistent with previous cloud models of the Jovian atmosphere \citep{sf02}.  

\section{Discussion}
\label{sect:discussion}

\subsection{Comparison to previous studies}

\cite{dyu16} previously performed fits to Pioneer and Cassini data of Jupiter using the DHG reflection law.  Instead of fitting $A_g \Psi$ to data, these authors directly fitted $P$ to the reflection coefficient; see Figures 1 and 2 of \cite{dyu16}.  Inferred values of the DHG fitting parameters are reported in Table 2 of \cite{dyu16}, but the associated uncertainties and posterior distributions are not reported as no formal uncertainty estimates were performed.  Figures 3 and 4 of \cite{dyu16} show comparisons of the phase curves of Jupiter to numerical calculations of $A_g \Psi$, but no fits of $A_g \Psi$ to data are shown.  \cite{mayorga16} and \cite{li18} fitted the phase curves with polynomials, but these fits do not allow $\omega$, $g_1$ and $g_2$ to be extracted.

\subsection{Coherent backscattering in the Jovian atmosphere?}

The physics behind the cuspy profiles of the Jovian phase curves remain unelucidated.  In Solar System bodies with solid surfaces and regolith, cuspy phase curves are caused by a combination of shadow hiding and coherent backscattering \citep{hapke98}.  The same effects have been observed for the rings of Saturn \citep{deau13}.  

Coherent backscattering is the phenomenon of multiply scattered light within a non-uniform medium interfering constructively to produce a brightness peak at zero phase angle \citep{hapke93,hapke02}.  It has been studied in a variety of contexts within physics (see \citealt{akk88} and references therein).

To explore the possibility that coherent backscattering is the cause of the cuspy nature of the Jovian phase curves near $\alpha=0^\circ$, we multiply $\rho$ by a simplified function that is given by equations (28) and (29) of \cite{hapke02},
\begin{equation}
1 + \frac{B_0}{1 + h^{-1} \tan{\left( \alpha/2 \right)}},
\end{equation}
where $B_0$ (magnitude of coherent backscattering) and $h$ (ratio of $\lambda/4\pi$ to transport mean free path) are two additional fitting parameters of the model.  In order to keep the number of parameters at five (including $\ln{\delta}$), we adopt the single Henyey-Greenstein reflection law ($f=1$, $g_1=g$).  Instead of fitting all 61 Jovian phase curves with the modification stated above, we focus on one example: 0.50 $\mu$m.  The fit to data, residuals and posterior distributions of $\omega$, $g$, $B_0$ and $h$ are shown in Figure \ref{fig:CB}.  Future studies should elucidate, from first principles, if coherent backscattering is a significant effect in gaseous atmospheres with large, irregular aerosols.

\subsection{Implications for exoplanets}

It is unknown if the large, irregular, polydisperse scatterers present in the Jovian atmosphere are also present in hot Jovian atmospheres.  If they are, a possible obstacle with detecting the cuspy profiles of reflected light phase curves at zero phase angle is that no information is available during secondary eclipse.  At the time of writing, the most robust detection of a reflected light phase curve is of the hot Jupiter Kepler-7b \citep{demory13}.  Figures 2 and 3 of \cite{demory13} show that the duration of the secondary eclipse corresponds to about 5\% of the orbit, which implies that information on any potentially cuspy profile is unavailable across $-9^\circ \lesssim \alpha \lesssim 9^\circ$.  Fits performed with data from $0^\circ \le \alpha \le 9^\circ$ excluded indicate that this does not alleviate the difficulty with fitting the single Henyey-Greenstein reflection law to the Jovian phase curves (not shown).  Cuspy profiles in the reflected light phase curves of gas-giant exoplanets remain to be detected.

There is a key difference between the interpretation of reflected light from Jupiter versus hot Jupiters.  Since its rotational and orbital periods are about 10 hours and 12 years, respectively, Jupiter is a fast rotator.  To lowest order, a fast rotator has longitudinal symmetry in terms of insolation and may be interpreted as a homogeneous sphere.  The fact that all 61 Cassini phase curves considered in the current study peak at $\alpha=0^\circ$ is strongly consistent with a homogeneous sphere.  By contrast, tidally locked hot Jupiters do not possess longitudinal symmetry and are interpreted as inhomogeneous spheres, which produce phase curves that do not peak at $\alpha=0^\circ$---an expectation that is borne out by both visible/optical photometric observations \citep{demory13,hu15,sh15} and general circulation models \citep{oreshenko16,par16,rr17}.

The current investigation suggests that precise, multi-wavelength phase curves that will be procured by the James Webb Space Telescope encode valuable information on the fundamental properties of clouds and hazes.  Retrieving for the single-scattering albedos and scattering asymmetry factors using an ab initio model will motivate further studies into the microphysics and chemistry of clouds and hazes in exoplanetary atmospheres.  In this context, the method described in the current Letter provides a crucial bridge between observations and simulations.

\begin{acknowledgments}
KH acknowledges financial support from the European Research Council (via a Consolidator Grant to KH; grant number 771620), the PlanetS National Center of Competence in Research (NCCR) and the Center for Space and Habitability (CSH), as well as a honorary professorship from Warwick University.  KH is grateful to his senior Ph.D student Chloe Fisher for teaching him how to use \texttt{emcee} and \texttt{corner} and to his postdoctoral research associate Brett Morris for advice on data fitting.
\end{acknowledgments}

\begin{table*}
\begin{center}
\caption{Double Henyey-Greenstein Parameters from Fits to Cassini Data}
\begin{tabular}{lccccc}
\hline
\hline
$\lambda$ ($\mu$m) & $\omega$ & $g_1$ & $g_2$ & $f$ & $\ln{\delta}$ \\
\hline
0.40 & $0.947^{+0.001}_{-0.001}$ & $0.381^{+0.006}_{-0.007}$ & $-0.658^{+0.005}_{-0.006}$ & $0.940^{+0.002}_{-0.002}$ & $-4.79^{+0.10}_{-0.10}$ \\
0.41 & $0.952^{+0.001}_{-0.001}$ & $0.359^{+0.007}_{-0.006}$ & $-0.643^{+0.005}_{-0.005}$ & $0.928^{+0.002}_{-0.002}$ & $-4.87^{+0.11}_{-0.11}$ \\
0.42 & $0.956^{+0.001}_{-0.001}$ & $0.332^{+0.006}_{-0.006}$ & $-0.628^{+0.005}_{-0.005}$ & $0.914^{+0.002}_{-0.003}$ & $-4.91^{+0.12}_{-0.12}$ \\
0.43 & $0.959^{+0.001}_{-0.001}$ & $0.298^{+0.007}_{-0.006}$ & $-0.613^{+0.006}_{-0.006}$ & $0.897^{+0.003}_{-0.003}$ & $-4.89^{+0.12}_{-0.12}$ \\
0.44 & $0.956^{+0.001}_{-0.001}$ & $0.248^{+0.007}_{-0.006}$ & $-0.593^{+0.006}_{-0.007}$ & $0.875^{+0.004}_{-0.004}$ & $-4.84^{+0.13}_{-0.12}$ \\
0.45 & $0.957^{+0.001}_{-0.001}$ & $0.205^{+0.006}_{-0.006}$ & $-0.585^{+0.007}_{-0.007}$ & $0.860^{+0.005}_{-0.005}$ & $-4.78^{+0.13}_{-0.12}$ \\
0.46 & $0.958^{+0.001}_{-0.001}$ & $0.183^{+0.007}_{-0.007}$ & $-0.585^{+0.007}_{-0.007}$ & $0.856^{+0.006}_{-0.006}$ & $-4.60^{+0.11}_{-0.11}$ \\
0.47 & $0.965^{+0.001}_{-0.001}$ & $0.206^{+0.008}_{-0.007}$ & $-0.597^{+0.007}_{-0.007}$ & $0.865^{+0.005}_{-0.006}$ & $-4.53^{+0.10}_{-0.10}$ \\
0.48 & $0.971^{+0.001}_{-0.001}$ & $0.226^{+0.008}_{-0.007}$ & $-0.609^{+0.008}_{-0.007}$ & $0.873^{+0.005}_{-0.005}$ & $-4.47^{+0.10}_{-0.10}$ \\
0.49 & $0.975^{+0.001}_{-0.001}$ & $0.246^{+0.008}_{-0.008}$ & $-0.621^{+0.008}_{-0.007}$ & $0.882^{+0.005}_{-0.005}$ & $-4.40^{+0.10}_{-0.10}$ \\
0.50 & $0.979^{+0.001}_{-0.001}$ & $0.266^{+0.008}_{-0.008}$ & $-0.631^{+0.008}_{-0.008}$ & $0.891^{+0.005}_{-0.005}$ & $-4.31^{+0.10}_{-0.10}$ \\
0.51 & $0.981^{+0.001}_{-0.001}$ & $0.282^{+0.008}_{-0.008}$ & $-0.639^{+0.009}_{-0.008}$ & $0.897^{+0.005}_{-0.005}$ & $-4.25^{+0.10}_{-0.09}$ \\
0.52 & $0.985^{+0.001}_{-0.001}$ & $0.303^{+0.008}_{-0.008}$ & $-0.655^{+0.008}_{-0.008}$ & $0.907^{+0.004}_{-0.005}$ & $-4.21^{+0.10}_{-0.09}$ \\
0.53 & $0.987^{+0.001}_{-0.001}$ & $0.315^{+0.009}_{-0.008}$ & $-0.666^{+0.008}_{-0.008}$ & $0.915^{+0.004}_{-0.004}$ & $-4.23^{+0.10}_{-0.10}$ \\
0.54 & $0.987^{+0.001}_{-0.001}$ & $0.317^{+0.009}_{-0.009}$ & $-0.669^{+0.008}_{-0.008}$ & $0.918^{+0.004}_{-0.004}$ & $-4.24^{+0.10}_{-0.10}$ \\
0.55 & $0.989^{+0.001}_{-0.001}$ & $0.337^{+0.010}_{-0.010}$ & $-0.685^{+0.008}_{-0.009}$ & $0.927^{+0.004}_{-0.004}$ & $-4.13^{+0.10}_{-0.10}$ \\
0.56 & $0.993^{+0.001}_{-0.001}$ & $0.364^{+0.012}_{-0.012}$ & $-0.714^{+0.011}_{-0.010}$ & $0.941^{+0.004}_{-0.004}$ & $-3.86^{+0.09}_{-0.09}$ \\
0.57 & $0.993^{+0.001}_{-0.001}$ & $0.366^{+0.013}_{-0.012}$ & $-0.724^{+0.011}_{-0.011}$ & $0.947^{+0.004}_{-0.004}$ & $-3.83^{+0.10}_{-0.10}$ \\
0.58 & $0.993^{+0.001}_{-0.001}$ & $0.372^{+0.011}_{-0.010}$ & $-0.729^{+0.010}_{-0.011}$ & $0.949^{+0.004}_{-0.004}$ & $-3.84^{+0.09}_{-0.09}$ \\
0.59 & $0.993^{+0.001}_{-0.001}$ & $0.374^{+0.010}_{-0.010}$ & $-0.731^{+0.010}_{-0.010}$ & $0.949^{+0.003}_{-0.004}$ & $-3.83^{+0.08}_{-0.07}$ \\
0.60 & $0.992^{+0.001}_{-0.001}$ & $0.366^{+0.008}_{-0.008}$ & $-0.723^{+0.009}_{-0.009}$ & $0.947^{+0.003}_{-0.003}$ & $-3.88^{+0.06}_{-0.06}$ \\
0.61 & $0.997^{+0.001}_{-0.001}$ & $0.430^{+0.017}_{-0.017}$ & $-0.861^{+0.032}_{-0.030}$ & $0.991^{+0.003}_{-0.004}$ & $-2.55^{+0.06}_{-0.06}$ \\
0.62 & $0.990^{+0.001}_{-0.001}$ & $0.435^{+0.018}_{-0.018}$ & $-0.870^{+0.056}_{-0.039}$ & $0.995^{+0.002}_{-0.003}$ & $-2.37^{+0.07}_{-0.06}$ \\
0.63 & $0.996^{+0.001}_{-0.001}$ & $0.424^{+0.014}_{-0.014}$ & $-0.826^{+0.022}_{-0.021}$ & $0.984^{+0.003}_{-0.004}$ & $-2.94^{+0.06}_{-0.06}$ \\
0.64 & $0.993^{+0.001}_{-0.001}$ & $0.425^{+0.011}_{-0.010}$ & $-0.764^{+0.013}_{-0.013}$ & $0.966^{+0.003}_{-0.003}$ & $-3.58^{+0.07}_{-0.07}$ \\
0.65 & $0.993^{+0.001}_{-0.001}$ & $0.453^{+0.014}_{-0.014}$ & $-0.801^{+0.020}_{-0.019}$ & $0.980^{+0.004}_{-0.004}$ & $-3.16^{+0.10}_{-0.10}$ \\
0.66 & $0.995^{+0.001}_{-0.001}$ & $0.449^{+0.016}_{-0.016}$ & $-0.837^{+0.028}_{-0.024}$ & $0.988^{+0.003}_{-0.004}$ & $-2.77^{+0.07}_{-0.07}$ \\
0.67 & $0.995^{+0.001}_{-0.001}$ & $0.450^{+0.016}_{-0.016}$ & $-0.854^{+0.033}_{-0.028}$ & $0.991^{+0.003}_{-0.004}$ & $-2.60^{+0.07}_{-0.07}$ \\
0.68 & $0.997^{+0.001}_{-0.001}$ & $0.446^{+0.018}_{-0.018}$ & $-0.868^{+0.048}_{-0.037}$ & $0.994^{+0.003}_{-0.004}$ & $-2.35^{+0.07}_{-0.06}$ \\
0.69 & $0.998^{+0.001}_{-0.001}$ & $0.443^{+0.020}_{-0.019}$ & $-0.864^{+0.084}_{-0.046}$ & $0.995^{+0.003}_{-0.004}$ & $-2.24^{+0.06}_{-0.06}$ \\
0.70 & $0.993^{+0.001}_{-0.001}$ & $0.443^{+0.018}_{-0.018}$ & $-0.867^{+0.043}_{-0.034}$ & $0.994^{+0.003}_{-0.004}$ & $-2.42^{+0.06}_{-0.06}$ \\
0.71 & $0.996^{+0.001}_{-0.001}$ & $0.442^{+0.019}_{-0.019}$ & $-0.858^{+0.155}_{-0.050}$ & $0.996^{+0.002}_{-0.004}$ & $-2.22^{+0.06}_{-0.06}$ \\
0.72 & $0.965^{+0.001}_{-0.001}$ & $0.433^{+0.010}_{-0.010}$ & $-0.802^{+0.016}_{-0.015}$ & $0.985^{+0.002}_{-0.002}$ & $-3.39^{+0.07}_{-0.07}$ \\
0.73 & $0.892^{+0.001}_{-0.001}$ & $0.340^{+0.005}_{-0.004}$ & $-0.614^{+0.006}_{-0.006}$ & $0.937^{+0.002}_{-0.002}$ & $-4.79^{+0.09}_{-0.09}$ \\
0.74 & $0.995^{+0.001}_{-0.001}$ & $0.446^{+0.019}_{-0.020}$ & $-0.834^{+0.371}_{-0.067}$ & $0.996^{+0.002}_{-0.004}$ & $-2.24^{+0.07}_{-0.06}$ \\
0.75 & $0.996^{+0.001}_{-0.001}$ & $0.450^{+0.021}_{-0.020}$ & $-0.827^{+0.482}_{-0.076}$ & $0.996^{+0.002}_{-0.004}$ & $-2.25^{+0.07}_{-0.07}$ \\
0.76 & $0.991^{+0.001}_{-0.001}$ & $0.451^{+0.018}_{-0.016}$ & $-0.855^{+0.042}_{-0.038}$ & $0.994^{+0.003}_{-0.003}$ & $-2.56^{+0.08}_{-0.08}$ \\
0.77 & $0.991^{+0.001}_{-0.001}$ & $0.448^{+0.017}_{-0.016}$ & $-0.857^{+0.043}_{-0.037}$ & $0.994^{+0.003}_{-0.003}$ & $-2.54^{+0.08}_{-0.08}$ \\
0.78 & $0.979^{+0.001}_{-0.001}$ & $0.455^{+0.014}_{-0.014}$ & $-0.817^{+0.023}_{-0.022}$ & $0.988^{+0.002}_{-0.003}$ & $-3.12^{+0.09}_{-0.09}$ \\
0.79 & $0.958^{+0.001}_{-0.001}$ & $0.466^{+0.011}_{-0.011}$ & $-0.734^{+0.010}_{-0.010}$ & $0.974^{+0.002}_{-0.002}$ & $-4.30^{+0.11}_{-0.12}$ \\
0.80 & $0.965^{+0.001}_{-0.001}$ & $0.457^{+0.011}_{-0.011}$ & $-0.767^{+0.015}_{-0.014}$ & $0.980^{+0.002}_{-0.002}$ & $-3.77^{+0.10}_{-0.10}$ \\
0.81 & $0.986^{+0.001}_{-0.001}$ & $0.447^{+0.016}_{-0.017}$ & $-0.859^{+0.036}_{-0.032}$ & $0.994^{+0.002}_{-0.003}$ & $-2.62^{+0.07}_{-0.07}$ \\
0.82 & $0.993^{+0.001}_{-0.001}$ & $0.445^{+0.019}_{-0.019}$ & $-0.860^{+0.115}_{-0.049}$ & $0.996^{+0.002}_{-0.003}$ & $-2.28^{+0.07}_{-0.06}$ \\
0.83 & $0.995^{+0.001}_{-0.001}$ & $0.445^{+0.019}_{-0.020}$ & $-0.819^{+0.510}_{-0.086}$ & $0.997^{+0.002}_{-0.004}$ & $-2.15^{+0.07}_{-0.06}$ \\
\hline
\hline
\end{tabular}\\
Note: Uncertainties are stated for $1\sigma$ (16th, 50th and 84th percentile). \\
\end{center}
\end{table*}

\begin{table*}
\tablenum{1}
\begin{center}
\caption{Double Henyey-Greenstein Parameters from Fits to Cassini Data (continued)}
\begin{tabular}{lccccc}
\hline
\hline
$\lambda$ ($\mu$m) & $\omega$ & $g_1$ & $g_2$ & $f$ & $\ln{\delta}$ \\
\hline
0.84 & $0.982^{+0.001}_{-0.001}$ & $0.443^{+0.015}_{-0.015}$ & $-0.858^{+0.034}_{-0.030}$ & $0.993^{+0.002}_{-0.003}$ & $-2.66^{+0.07}_{-0.06}$ \\
0.85 & $0.988^{+0.001}_{-0.001}$ & $0.442^{+0.017}_{-0.017}$ & $-0.868^{+0.051}_{-0.038}$ & $0.995^{+0.002}_{-0.003}$ & $-2.39^{+0.07}_{-0.06}$ \\
0.86 & $0.868^{+0.001}_{-0.001}$ & $0.321^{+0.005}_{-0.005}$ & $-0.582^{+0.007}_{-0.007}$ & $0.927^{+0.003}_{-0.003}$ & $-4.56^{+0.07}_{-0.07}$ \\
0.87 & $0.906^{+0.001}_{-0.001}$ & $0.377^{+0.005}_{-0.005}$ & $-0.648^{+0.006}_{-0.006}$ & $0.951^{+0.002}_{-0.002}$ & $-4.62^{+0.07}_{-0.07}$ \\
0.88 & $0.760^{+0.001}_{-0.001}$ & $0.252^{+0.006}_{-0.006}$ & $-0.476^{+0.011}_{-0.011}$ & $0.877^{+0.006}_{-0.007}$ & $-4.18^{+0.07}_{-0.06}$ \\
0.89 & $0.304^{+0.001}_{-0.001}$ & $0.125^{+0.012}_{-0.010}$ & $-0.373^{+0.021}_{-0.021}$ & $0.836^{+0.018}_{-0.022}$ & $-4.06^{+0.10}_{-0.08}$ \\
0.90 & $0.473^{+0.001}_{-0.001}$ & $0.175^{+0.011}_{-0.010}$ & $-0.368^{+0.018}_{-0.019}$ & $0.814^{+0.018}_{-0.019}$ & $-4.30^{+0.11}_{-0.11}$ \\
0.91 & $0.964^{+0.001}_{-0.001}$ & $0.439^{+0.012}_{-0.013}$ & $-0.829^{+0.025}_{-0.024}$ & $0.991^{+0.002}_{-0.003}$ & $-2.98^{+0.06}_{-0.06}$ \\
0.92 & $0.980^{+0.001}_{-0.001}$ & $0.441^{+0.016}_{-0.017}$ & $-0.861^{+0.046}_{-0.037}$ & $0.995^{+0.002}_{-0.003}$ & $-2.48^{+0.07}_{-0.06}$ \\
0.93 & $0.982^{+0.001}_{-0.001}$ & $0.441^{+0.018}_{-0.016}$ & $-0.858^{+0.062}_{-0.045}$ & $0.996^{+0.002}_{-0.003}$ & $-2.40^{+0.06}_{-0.06}$ \\
0.94 & $0.991^{+0.001}_{-0.001}$ & $0.440^{+0.021}_{-0.019}$ & $-0.583^{+0.453}_{-0.285}$ & $0.997^{+0.002}_{-0.004}$ & $-2.11^{+0.06}_{-0.06}$ \\
0.95 & $0.989^{+0.001}_{-0.001}$ & $0.443^{+0.019}_{-0.020}$ & $-0.712^{+0.528}_{-0.168}$ & $0.997^{+0.002}_{-0.004}$ & $-2.15^{+0.06}_{-0.06}$ \\
0.96 & $0.975^{+0.001}_{-0.001}$ & $0.442^{+0.016}_{-0.016}$ & $-0.850^{+0.047}_{-0.036}$ & $0.995^{+0.002}_{-0.003}$ & $-2.54^{+0.06}_{-0.06}$ \\
0.97 & $0.785^{+0.001}_{-0.001}$ & $0.285^{+0.005}_{-0.004}$ & $-0.498^{+0.007}_{-0.007}$ & $0.898^{+0.003}_{-0.003}$ & $-4.68^{+0.08}_{-0.08}$ \\
0.98 & $0.757^{+0.001}_{-0.001}$ & $0.262^{+0.006}_{-0.006}$ & $-0.483^{+0.008}_{-0.009}$ & $0.893^{+0.004}_{-0.004}$ & $-4.60^{+0.09}_{-0.09}$ \\
0.99 & $0.419^{+0.002}_{-0.002}$ & $0.083^{+0.010}_{-0.009}$ & $-0.502^{+0.022}_{-0.021}$ & $0.925^{+0.010}_{-0.012}$ & $-4.73^{+0.13}_{-0.14}$ \\
1.00 & $0.571^{+0.002}_{-0.002}$ & $0.112^{+0.012}_{-0.010}$ & $-0.513^{+0.018}_{-0.017}$ & $0.921^{+0.008}_{-0.010}$ & $-4.61^{+0.11}_{-0.11}$ \\
\hline
\hline
\end{tabular}\\
Note: Uncertainties are stated for $1\sigma$ (16th, 50th and 84th percentile). \\
\end{center}
\end{table*}

\label{lastpage}

\end{document}